\documentclass[fleqn,usenatbib]{mnras}

\usepackage{newtxtext,newtxmath}

\usepackage[T1]{fontenc}

\DeclareRobustCommand{\VAN}[3]{#2}
\let\VANthebibliography\thebibliography
\def\thebibliography{\DeclareRobustCommand{\VAN}[3]{##3}\VANthebibliography}

\usepackage{color}
\usepackage{graphicx}	
\usepackage{amsmath}	

\title[]{Quasars with Flare/Eclipse-like Variability Identified in ZTF}

\author[Zheng et al.]{
Zhiyuan Zheng,$^{1}$
Yong Shi,$^{1,2}$\thanks{E-mail: yong@nju.edu.cn}
Shuowen Jin,$^{3,4,5}$
H. Dannerbauer,$^{5,6}$
Qiusheng Gu,$^{1,2}$
Xin Li, $^{1}$
Xiaoling Yu $^{7}$
\\
$^{1}$School of Astronomy and Space Science, Nanjing University, Nanjing 210093, People's Republic of China\\
$^{2}$Key Laboratory of Modern Astronomy and Astrophysics (Nanjing University), Ministry of Education, Nanjing 210093, People's Republic of China\\
$^{3}$Cosmic Dawn Center (DAWN), Copenhagen, Denmark\\
$^{4}$DTU Space, Technical University of Denmark, Elektrovej 327, 2800 Kgs. Lyngby, Denmark\\
$^{5}$Instituto de Astrof\'{i}sica de Canarias (IAC), E-38205, La Laguna, Tenerife, Spain\\
$^{6}$Universidad de La Laguna, Dpto. Astrof\'{i}sica, E-38206, La Laguna, Tenerife, Spain\\
$^{7}$College of Physics and Electronic Engineering, Qujing Normal University, Qujing 655011, P.R. China\\
}

\date{Accepted XXX. Received YYY; in original form ZZZ}

\pubyear{2024}

\begin{document}
\label{firstpage}
\pagerange{\pageref{firstpage}--\pageref{lastpage}}
\maketitle

\begin{abstract}

Active galactic nuclei (AGNs) are known to exhibit optical/UV variability and most of them can be well modeled by the damped random walks. Physical processes that are not related to the accretion disk, such as tidal disruption events (TDE) or moving foreground dusty clouds, can cause flare-like and eclipse-like features in the optical light curve. Both long-term and high-cadence monitoring are needed to identify such features. By combining the Sloan Digital Sky Survey (SDSS), Panoramic Survey Telescope, and Rapid Response System (Pan-STARRS) with the Zwicky Transient Facility (ZTF) survey, we are able to identify a rare sample (11) out of the SDSS quasar catalog ($\sim 83,000$). These quasars exhibit more or less constant brightness but show rapid optical variation in the ZTF DR2 epochs. To investigate the possible origins of these flare/eclipse-like variabilities, we propose the second epoch spectroscopic observations with the Gran Telescopio CANARIAS (GTC). We find that the change in accretion rate plays a significant role in these quasar variabilities. Among them, we identify two Changing-Look Active Galactic Nuclei (CL-AGN) candidates: SDSS J1427+2930 and SDSS J1420+3757. The luminosity change of the former may be caused by the enhanced SMBH's accretion or the tidal disruption event, while the latter is more related to the change in the accretion rate.

\end{abstract}




\section{Introduction}

Supermassive black holes (SMBHs) are located at the centers of almost all massive galaxies, and they grow their mass by accreting the surrounding material \citep{2013ARA&A..51..511K}. The accreting process emits huge radiation across the entire electromagnetic spectra. These active galactic nuclei (AGNs) affect the formation and evolution of galaxies through their radiative and mechanical feedback \citep{2015ARA&A..53..115K}. AGNs are empirically classified as type 1 and type 2 based on their observational features in optical/ultraviolet (UV) spectra. Type 1 AGNs show both broad (i.e. Full Width at Half Maximum, FWHM $>1200\ \rm km\;s^{-1}$) and narrow (FWHM $<1200\ \rm km\;s^{-1}$) emission lines, while those only displaying narrow emission lines are referred to as type 2 AGNs. The unification model provides a reasonable explanation for the apparent AGN types in observations \citep{1993ARA&A..31..473A,2015ARA&A..53..365N}, which states that the different AGN types are observed through various viewing angles to the same structures. Despite its huge success in explaining the observed AGN spectra, it also faces some challenges, such as the lack of intrinsic broad line emission in some AGNs \citep{2010ApJ...714..115S, 2019MNRAS.488L...1B}, or the existence of changing-look AGNs \citep[CL-AGNs,][]{2016MNRAS.457..389M, 2022arXiv221105132R}.

AGNs are known to be highly variable in electromagnetic radiation during their accretion phase   \citep{1997ARA&A..35..445U, 2001sac..conf....3P, 2005MNRAS.359..345U, 2023MNRAS.522..767J}. The AGN variability provides a powerful tool to investigate the structures near the SMBH. In the past decades, based on the large area multi-epoch optical survey including the Sloan Digital Sky Survey \citep[SDSS,][]{2000AJ....120.1579Y}, Panoramic Survey Telescope and Rapid Response System \citep[Pan-STARRS,][]{2010SPIE.7733E..0EK}, the Palomar Transient Factory \citep[PTF,][]{2009PASP..121.1395L} etc., large samples are built to characterize the optical AGN variability \citep{2004ApJ...601..692V, 2016MNRAS.457..389M, 2018ApJ...862..109Y, 2019ApJ...874....8M, 2022ApJ...933..180G}. In general, AGNs show stochastic moderate variabilities over years \citep{2002ApJ...581..197P, 2004ApJ...601..692V}. On the other hand, there is a growing number of dramatic luminosity changes of AGNs that undergo extreme brightening or dimming within months or years \citep{2018ApJ...854..160R, 2021ApJ...921...70S}. 

Those AGNs with dramatic luminosity changes have the potential to be recognized as CL-AGNs. These events can be further classified into two distinct types based on their origin: Changing-Obscuration AGNs (CO-AGNs) or Changing-State AGNs (CS-AGNs). Typically, CO-AGNs are usually identified in X-ray observations, where they exhibit huge variations in the column density ($N_{\rm H}$) \citep{2007MNRAS.377..607P, 2003MNRAS.342..422M, 2012MNRAS.421.1803M}. In contrast, CS-AGNs are discovered through optical/UV spectroscopic observations, where their broad emission lines appear or disappear over years \citep{1986ApJ...311..135C, 2014ApJ...796..134D, 2014ApJ...788...48S, 2015ApJ...800..144L, 2020MNRAS.491.4925G}. The short timescale variations of BLR in CS-AGNs challenge both the unification model and the traditional accretion disk model \citep{1973A&A....24..337S, 1999agnc.book.....K}. 

Such a unique AGN property can be attributed to several underlying physical mechanisms, including changing obscuration, instabilities in accretion disks, and major perturbations/eruptions in the accretion disk. Obscuration variation, which has been observed in X-ray observations, can be caused by gas or dust clouds traveling through the line of sight \citep{2002ApJ...571..234R, 2004MNRAS.351..169M, 2009ApJ...696..160R, 2022ApJ...930...53L} or by changes in the ionization state of the obscuring material associated with intrinsic AGN luminosity variation \citep{1989MNRAS.236..153Y, 2005ApJ...623L..93R}. Instabilities in accretion disks can be driven by various effects and can distort the light curve across a large range of time scales. The thermal instability is driven by the changes in opacity and radiation pressure \citep{1986ApJ...305...28L}, which leads to the temperature fluctuations that are well modeled by the damped random walks \citep{2009ApJ...698..895K, 2010ApJ...721.1014M}. Another is the interactions between the accretion flow and an inflated disk that cause the changes in magnetic torques and further, perturb the accretion disk \citep{2018ApJ...864...27S}. Direct perturbations also can trigger extreme changes in the light curve. Tidal disruption events (TDEs) can generate a unique shape in the optical/UV light curve ($\propto t^{-5/3}$) when the SMBH captures and destroys a nearby star \citep{2015MNRAS.452...69M, 2019ApJ...881..113C, 2021ARA&A..59...21G}. Recent simulations have investigated the possible shapes of optical/UV light curves when TDEs and AGNs co-exist \citep{2022MNRAS.514.4102M, 2024MNRAS.527.8103R}. Other rare conditions could disturb the accretion disk significantly, such as the interactions between disks in binary SMBHs \citep{2020A&A...643L...9W}, or between the disk and a recoiling SMBH \citep{2018ApJ...861...51K}. 

In spite of different mechanisms that drive AGN optical variability, those that are not related to the accretion rate, such as TDE and foreground obscuration, can cause some distinct features in the light curve, i.e., a more or less constant light curve superposed with a flare-like feature and an eclipse-like feature, respectively. To identify such a population, a long timeline with a high cadence is needed, which is unfortunately not available yet. But by combining SDSS, Pan-STARRS with the Zwicky Transient Facility \citep[ZTF,][]{2019PASP..131f8003B, 2019PASP..131a8003M}, we can partially achieve the requirement and identify candidates. To investigate the nature of these candidates, we propose the second epoch spectroscopic observations with the Optical System for Imaging and low-intermediate-Resolution Integrated Spectroscopy (OSIRIS) instrument on Gran Telescopio CANARIA (GTC). 

This paper is organized as follows, in \S~2, we introduce our observations with GTC and how we reduce the raw data. In \S~3, the basic methods we use to treat the spectra and light curves are listed. In \S~4, we show the statistical features of our whole sample. In \S~5, we discuss the possible origins of the flare/eclipse-like variability and 2 CL-AGN candidates. Finally, we conclude this work in \S~6. The cosmological model are assumed as: $H_0 = 67.4$   $\rm km\;s^{-1}\;Mpc^{-1}$, $\Omega_{\rm m} = 0.315$ and $\Omega_{\rm \Lambda} = 0.685 $ \citep{2020A&A...641A...6P}.


\section{Observations and data analysis}

\subsection{Observations and Sample Selection}

\begin{table}
	\centering
	\caption{(1) Object name. (2) Redshift from NED. (3) Total exposure time for GTC spectra. (4) Actual seeing during observations. (5) Observing time in MJD.}
	\label{tab:obs_info} 
	\begin{tabular}{lcccccr} 
 
		\hline
            \hline
            SDSS name $^{(1)}$ & z $^{(2)}$&  $t_{\rm exp}^{(3)} (\rm sec)$& seeing $^{(4)}$& MJD$ ^{(5)}$  \\
            \hline
            J161700.81+124332.2 & 0.35343 & 540  & 0.9 \arcsec & 58997.2  \\ 
            J141802.79+414935.2 & 1.04189 & 540  & 0.9 \arcsec & 59046.9  \\  
            J142746.39+293038.2 & 0.37362 & 540  & 1.2 \arcsec & 59055.0  \\  
            J144050.76+520445.9 & 0.31875 & 350  & 1.0 \arcsec & 59044.0  \\ 
            J142048.66+375755.4 & 0.78662 & 600  & 1.4 \arcsec & 59044.9  \\ 
            J145206.89+360226.0 & 2.97118 & 720  & 0.9 \arcsec & 59044.0  \\ 
            J160353.04+264515.9 & 1.33399 & 540  & 0.9 \arcsec & 58990.2  \\ 
            J141119.44+413802.3 & 1.12984 & 720  & 1.2 \arcsec & 59046.9  \\ 
            J160747.76+304418.6 & 0.46507 & 540  & 0.9 \arcsec & 58990.2  \\ 
            J172800.66+545302.7 & 0.24549 & 360  & 0.7 \arcsec & 58995.2  \\ 
            J212715.33-062041.6 & 0.70392 & 540  & 0.8 \arcsec & 58995.2  \\ 
		\hline
  
       \end{tabular}
       \footnotesize{NED: \url{https://ned.ipac.caltech.edu/Library/Distances/}}
\end{table}

\begin{figure*}
   \resizebox{\hsize}{!}
             {\includegraphics[width=\textwidth]{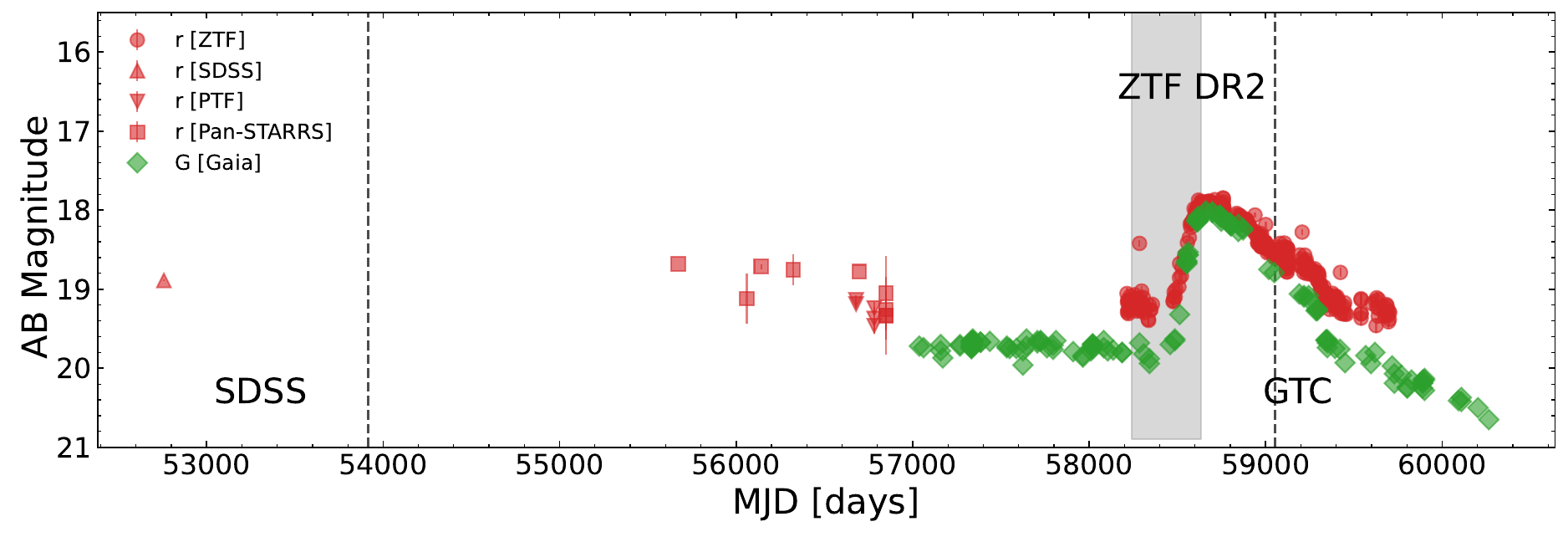}}
      \caption{Optical light curve of SDSS J1427+2930. The green diamonds mark the $g$ band light curve from Gaia, and the red triangles, squares, anti-triangles, and circles represent the $r$ band light curve from SDSS, Pan-STARRS, PTF, and ZTF, respectively. The grey shadowed region represents the ZTF DR2 epochs (March 2018 to June 2019), these quasars show rapid luminosity variation in this period. The two dashed vertical lines refer to when SDSS and GTC spectra were observed.}
         \label{fig:sampleSelect}
   \end{figure*}

To identify a flare-like or eclipse-like light curve, we first compiled a parent sample of hundreds of quasars exhibiting rapid luminosity variation by cross-matching the luminous SDSS quasars ($m_{\rm r-band} < 20$) \citep{2020ApJS..250....8L} with the ZTF Public Data Release 2 (DR2, March 2018 to June 2019, released on December 11, 2019). The rapid luminosity variation of quasars refers to their $r$-band magnitude undergoing a change larger than $0.35$ mag within a duration of less than half a year (at the rest-frame), as observed in the ZTF DR2 epochs. We then retrieve their optical light curves in early epochs from SDSS \citep{2019ApJS..240...23A} and Pan-STARRS1 \citep{2020ApJS..251....7F}. The random variation of quasars in early epochs is limited to less than 0.15 mag until the ZTF epoch. Subsequently, 19 quasars are selected that show recent significant luminosity changes while maintaining almost constant in their previous light curves. Figure~\ref{fig:sampleSelect} shows an example light curve of our sample. Eleven of 19 quasars are observed due to the constraints imposed by the all-weather mode of GTC observations. The new spectroscopic observation was done with GTC/OSIRIS (PID: GTC40-20A, PI: Shuowen Jin).

OSIRIS \citep{2000SPIE.4008..623C} is an imager and spectrograph onboard GTC. We took the new epoch long-slit spectra with the R1000B and R1000R grism at a slit width of $0.8 \arcsec$. The R1000B grism covers the wavelength range from 3630 to 7500 \AA\ with a spectral resolution of about 1000, while the R1000R grism covers the wavelength range from 5100 to 10,000 \AA\ with a spectral resolution of about 1100. The average exposure time on our science objects is about 540 secs. The typical seeing is around $0.9 \arcsec$.

\subsection{Data Reduction}

We adopt the \texttt{Python} program \texttt{PypeIt} \footnote{\url{https://pypeit.readthedocs.io/en/latest/}} \citep{2020zndo...3743493P, 2020JOSS....5.2308P} to reduce the raw data obtained from GTC OSIRIS. \texttt{PypeIt} is a highly configurable tool to reduce standard slit-imaging spectrographs, which is also adapted to the GTC-OSIRIS long-slit spectrograph. We follow the standard procedure to reduce the data. First, the science frames are calibrated for the bias and flat frames. Then the arc lamp emissions are used for wavelength calibrations. The flux correction uses the spectra of standard stars from the GTC preferred spectrophotometric standard star list for the optical wavelength range. We also check the flux calibration results by calibrating the standards by themselves and comparing the corrected spectra with the archived data. After combining the single exposures, the telluric correction is done with the \texttt{qso} mode for our objects. Due to the lack of an atmospheric model for GTC, the Maunakea model is adopted as an approximation. Subsequently, the telluric corrected spectra of two grisms are combined by the customized \texttt{Python} program based on flux conservation. Finally, the absolute flux calibration of spectra is based on ZTF $g$ and $r$ band photometry. The spectra are scaled by a constant to ensure their broad-band photometry is consistent with the ZTF photometry obtained at a similar observing time.

\section{Methods}

\begin{table*}
	\centering
	\caption{Physical properties of these quasars. Object name. (a) Viral SMBH mass estimated from the broad-line FWHM and flux. (b, c) Equivalent width. The EW of (b) and (c) are used for the SDSS and GTC epoch, respectively. (d, e) The continuum intensity. (f, g) The AGN bolometric luminosity. (h, i) Eddington ratio. (j, k) time-lag between $r$ and W1, W2 band. The different epochs of measurements are marked in the upper right corner. The different broad-line are utilized to measure these properties according to various $z$, detailed descriptions are seen in \S~4.2 and \S~4.3.}
	\label{tab:object_prop}
        \resizebox{1.0\textwidth}{!}{
	\begin{tabular}{lccccccccr}
 
		\hline
            \hline
            SDSS name & $^{(a)} {\rm Log}\ M_{\rm BH} (\rm M_{\odot})$ & $ ^{(b)} {\rm Log}\ W_{\lambda, {\rm br}}^{\rm SDSS}$ (\AA) & $ ^{(d)} {\rm Log}\ L_{\rm conti}^{\rm SDSS} (\rm erg\;s^{-1}$) & $ ^{(f)} {\rm Log}\ L_{\rm bol}^{\rm SDSS} (\rm erg\;s^{-1})$ & $ ^{(h)} \rm Log\ \eta^{\rm SDSS}$ & $^{(j)} {\rm \Delta t}_{W1-r}$ (days) & $ ^{(k)} {\rm \Delta t}_{W2-r}$ (days)  \\

            &  & $ ^{(c)} {\rm Log}\ W_{\lambda, {\rm br}}^{\rm GTC}$ (\AA) & $ ^{(e)} {\rm Log}\ L_{\rm conti}^{\rm GTC} (\rm erg\;s^{-1}$) & $ ^{(g)}  {\rm Log}\ L_{\rm bol}^{\rm GTC} (\rm erg\;s^{-1})$ & $ ^ {(i)} \rm Log\ \eta^{\rm GTC}$  &  \\
            \hline
            
            J161700.81+124332.2 & $ 7.96 \pm 0.13$ & $2.18 \pm 0.02$ & $43.96 \pm 0.02$ & $ 44.92 \pm 0.02$ & $-1.13 \pm 0.12$ & $629.0^{+287.3}_{-194.5}$ & $523.0^{+774.3}_{-290.0}$ \\
            &  & $1.85 \pm 0.04$ & $44.15 \pm 0.01$ & $ 45.12 \pm 0.01$ & $-0.94 \pm 0.12$ &  \\
        
            J141802.79+414935.2 & $ 9.42 \pm 0.19$ & $1.89 \pm 0.02$ & $45.28 \pm 0.01$ & $ 46.00 \pm 0.01$ & $ -1.51 \pm 0.19$ & $418.0^{+544.0}_{-309.3}$ & $481.0^{+663.5}_{-279.4}$ \\  
            &  & $1.67 \pm 0.01$ & $45.49 \pm 0.01$ & $ 46.20 \pm 0.01 $ & $ -1.30 \pm 0.18$ &  \\
            
            J142746.39+293038.2 & $ 8.05 \pm 0.22$ & $1.40 \pm 0.01$ & $44.19 \pm 0.01$ & $ 45.16 \pm 0.01 $ & $ -0.97 \pm 0.21 $ & $314.0^{+77.3}_{-120.6}$ & $409.0^{+128.2}_{-95.3}$ \\ 
            &  & $1.85 \pm 0.11$ & $44.03 \pm 0.01$ & $ 45.00 \pm 0.01 $ & $ -1.13 \pm 0.21 $  &  \\
            
            J144050.76+520445.9 & $ 8.25  \pm 0.11$ & $2.19 \pm 0.02$ & $44.40 \pm 0.01$ & $ 45.37 \pm 0.01 $ & $-0.97 \pm 0.11$ & $504.0^{+203.7}_{-151.6}$ & $504.0^{+216.9}_{-240.4}$ \\ 
            &  & $1.91 \pm 0.01$ & $44.26 \pm 0.01$ & $ 45.23 \pm 0.01 $ & $ -1.11 \pm 0.11 $  &  \\
            
            J142048.66+375755.4 & $ 9.14 \pm 0.17$ & $1.90 \pm 0.01$ & $45.12 \pm 0.01$ & $ 45.84 \pm 0.01 $ & $ -1.29 \pm 0.16 $ & $365.0^{+336.2}_{-329.7}$ & $421.0^{+335.6}_{-295.3}$ \\ 
            &  & $2.13 \pm 0.04$ & $44.01 \pm 0.02$ & $44.72 \pm 0.02 $ & $ -2.41 \pm 0.16 $   &  \\
            
            J145206.89+360226.0 & $ 9.34 \pm 0.08$ & $1.37 \pm 0.01$ & $46.67 \pm 0.01$ & $ 47.25 \pm 0.01 $ & $ -0.15 \pm 0.07 $ &  $417.0^{+610.1}_{-261.6}$ & $226.0^{+696.3}_{-508.6}$ \\ 
            &  & $1.49 \pm 0.01$ & $46.25 \pm 0.01$ & $ 46.83 \pm 0.01 $ & $ -0.58 \pm 0.07 $  &  \\
            
            J160353.04+264515.9 & $ 9.21 \pm 0.14$ & $1.41 \pm 0.02$ & $46.12 \pm 0.01$ & $ 46.84 \pm 0.01 $ & $ -0.46 \pm 0.14 $ & $313.0^{+935.5}_{-524.6}$ & $576.0^{+1010.0}_{-279.4}$ \\
            &  & $1.61 \pm 0.01$ & $45.89 \pm 0.01$ & $ 46.60 \pm 0.02 $ & $ -0.71 \pm 0.14 $ &  \\
            
            J141119.44+413802.3 & $ 9.49 \pm 0.06$ & $1.78 \pm 0.02$ & $45.41\pm 0.01$ & $ 46.12 \pm 0.01 $ & $ -1.45 \pm 0.06$ & $326.0^{+967.5}_{-391.2}$ & $408.0^{+989.2}_{-322.4}$ \\ 
            &  & $2.09 \pm 0.01$ & $44.95 \pm 0.01$ & $ 45.67 \pm 0.01 $ & $ -1.90 \pm 0.06 $  &  \\
            
            J160747.76+304418.6 & $ 8.19 \pm 0.36$ & $1.79 \pm 0.02$ & $44.51 \pm 0.01$ & $ 45.47 \pm 0.01 $ & $ -0.58 \pm 0.36$ & $402.0^{+250.0}_{-242.4}$ & $397.0^{+213.0}_{-195.9}$ \\ 
            &  & $1.41 \pm 0.12$ & $44.52 \pm 0.01$ & $ 45.49 \pm 0.02 $ & $ -0.58 \pm 0.36 $  &  \\
            
            J172800.66+545302.7 & $ 7.83 \pm 0.09$ & $1.68 \pm 0.01$ & $44.11 \pm 0.01$ & $ 45.08 \pm 0.01 $ & $ -0.84 \pm 0.09 $ & $186.0^{+53.1}_{-97.4}$ & $189.0^{+235.7}_{-140.3}$ \\
            &  & $1.67 \pm 0.05$ & $43.95 \pm 0.01$ & $ 44.92 \pm 0.01 $ & $ -1.01 \pm 0.09 $  &  \\
            
            J212715.33-062041.6 & $ 9.36 \pm 0.15$ & $2.02 \pm 0.02$ & $45.33 \pm 0.01$ & $ 46.30 \pm 0.01 $ & $ -1.07 \pm 0.14 $ & $744.0^{+470.9}_{-261.1}$ & $566.0^{+763.6}_{-435.8}$ \\
            &  & $2.17 \pm 0.01$ & $45.14 \pm 0.01$ & $ 46.10 \pm 0.01 $ & $ -1.27 \pm 0.15 $  &  \\
            
		\hline
       \end{tabular}}
       
\end{table*}

\begin{figure}
   \centering
   \includegraphics[width=\columnwidth]{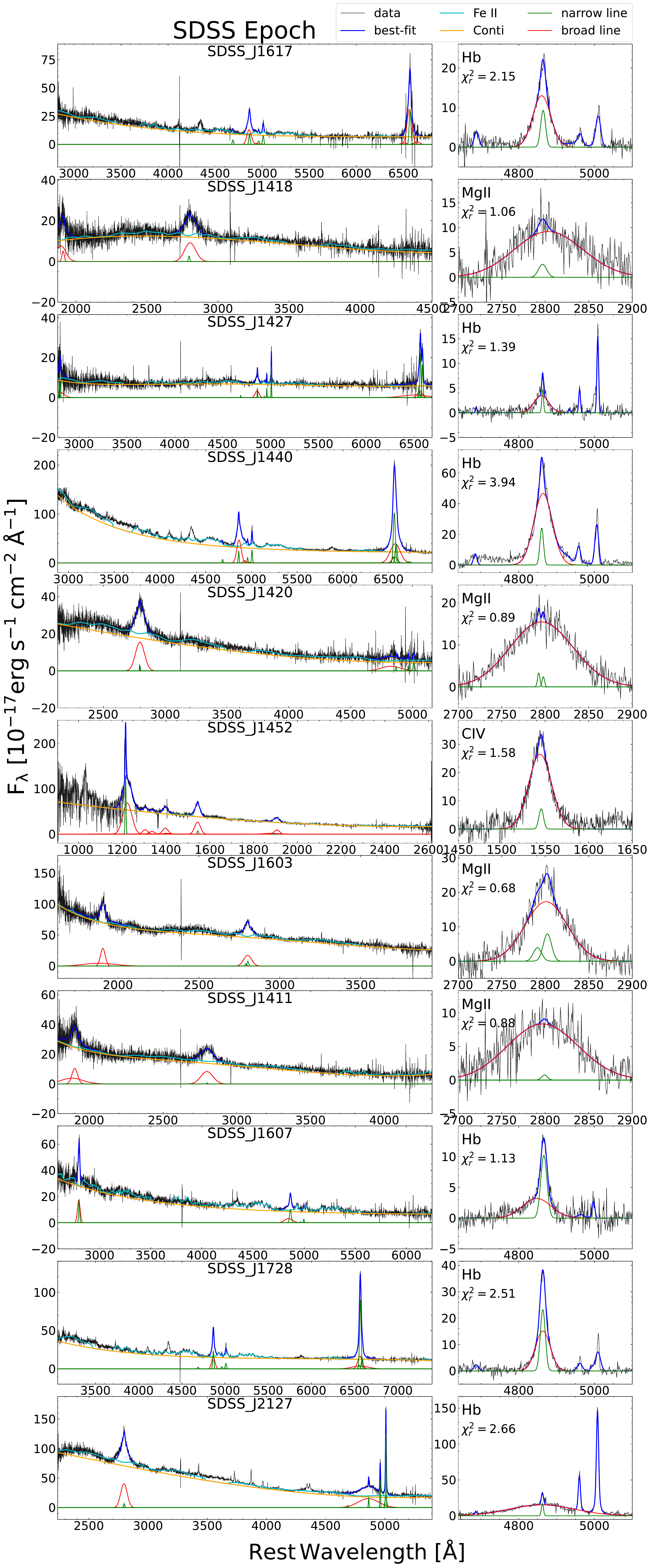}
      \caption{Spectra fitting results of all 11 quasars at SDSS epoch by \texttt{pyQSOFiT}. The right column shows the broad emission line fitting results. It shows different broad emission lines at different redshift: $z < 0.7$: $\rm H \beta $, $0.7 < z < 1.9$: \ion{Mg}{ii}, $z > 1.9$: \ion{C}{iv}.}
         \label{fig:all_sdss_spec}
   \end{figure}

\begin{figure}
   \centering
   \includegraphics[width=\columnwidth]{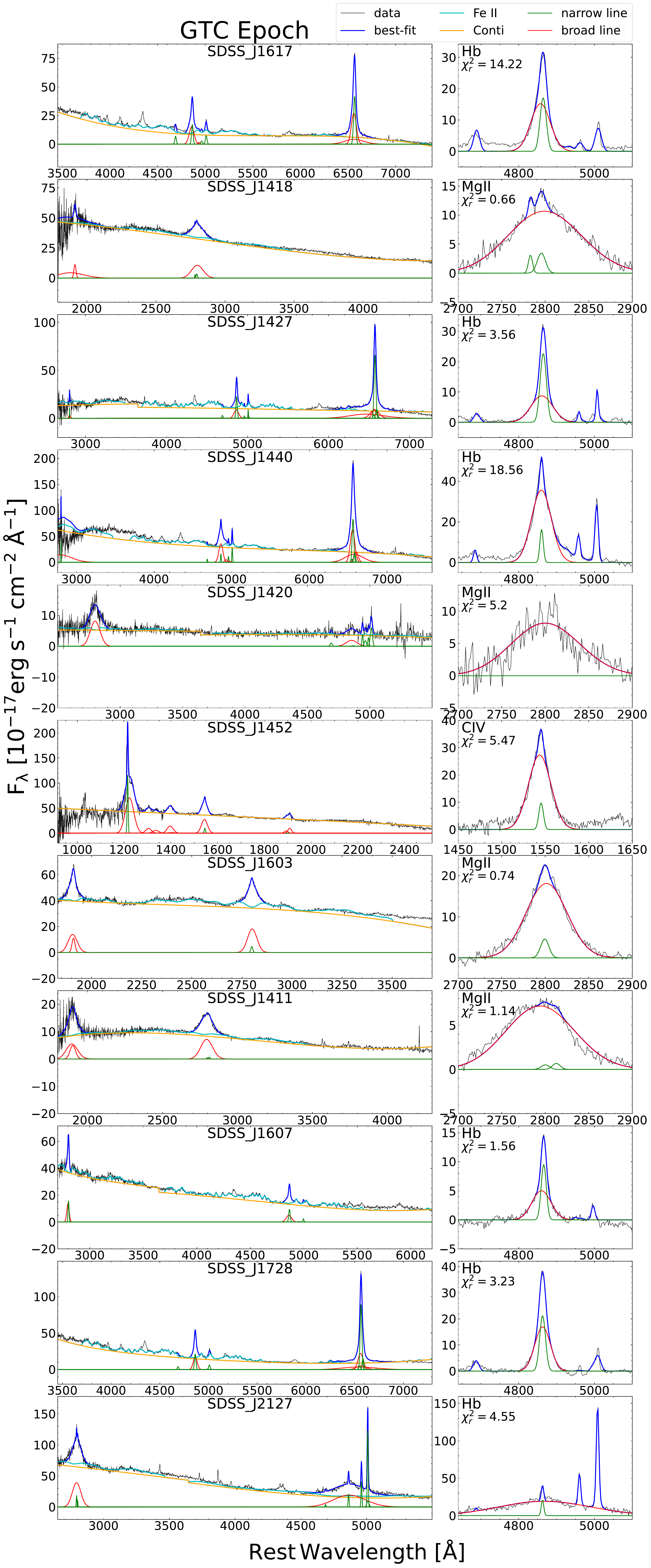}
      \caption{Spectra fitting results of all 11 quasars at GTC epoch by \texttt{pyQSOFiT}, symbols are similar as Figure~\ref{fig:all_sdss_spec}}
         \label{fig:all_gtc_spec}
   \end{figure}

\begin{figure}
   \centering
   \includegraphics[width=\columnwidth]{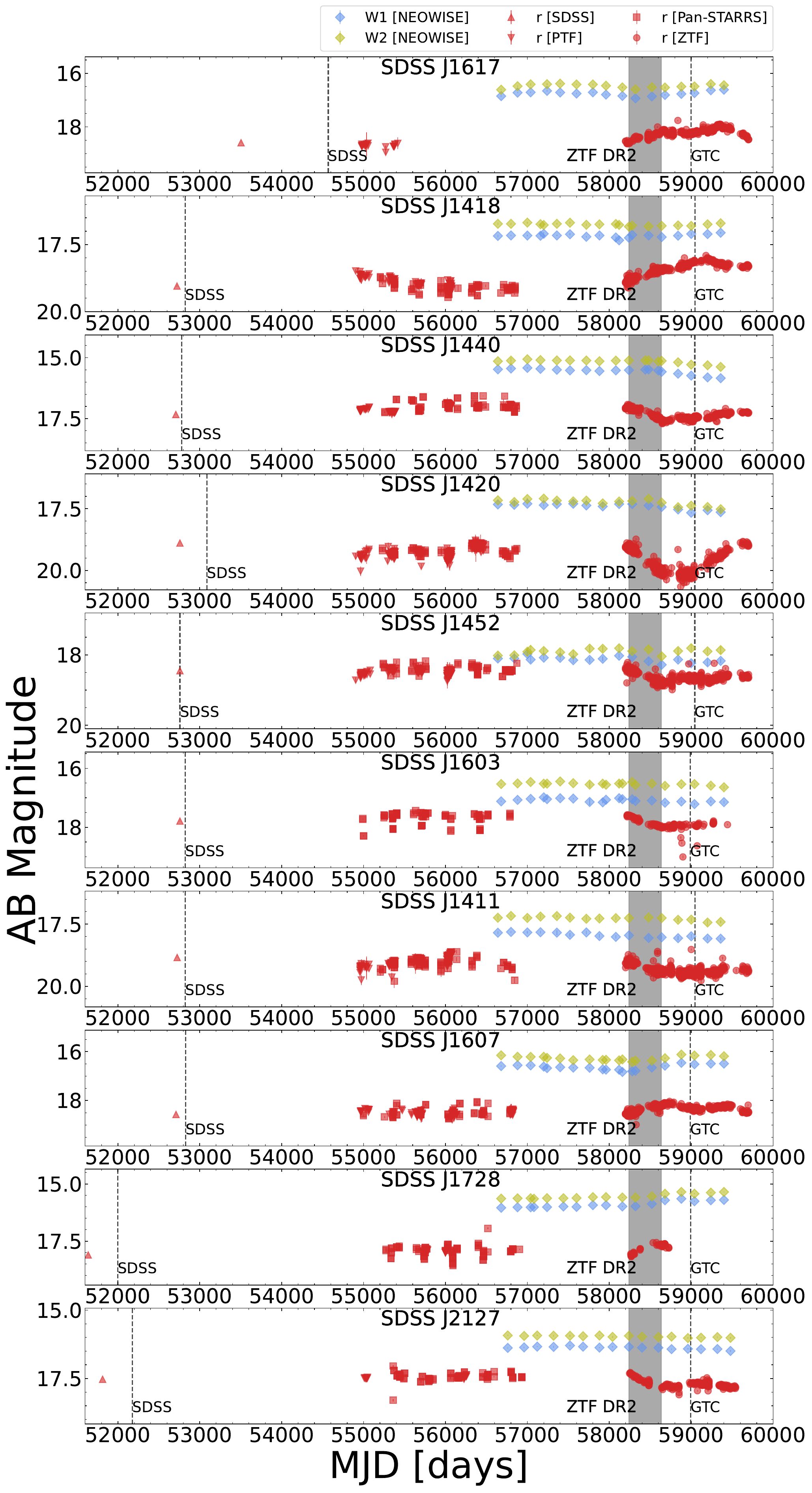}
      \caption{Optical light curves of other ten quasars, symbols are similar as Figure~\ref{fig:sampleSelect}.}
         \label{fig:all_lightCurve}
   \end{figure}


\subsection{Spectra Fitting}

Besides the GTC spectra obtained at the MJD of about 59000, we also collected the SDSS spectra at around MJD 53000. Combining the two epoch spectra, we can further study the origin of our quasars. We adopted the \texttt{PyQSOFit} \citep{2018ascl.soft09008G, 2019MNRAS.482.3288G} to decompose the spectra. This program uses a $\chi^2$-based method to fit the spectra with spectral models and templates. Briefly, the quasar spectra are decomposed into the power-law continuum plus \ion{Fe}{ii} templates \citep{1992ApJS...80..109B, 2001ApJS..134....1V}, along with a series of Gaussian profiles for emission lines. For each emission line, sometimes multiple Gaussian components are needed in order to fully fit the line due to the presence of broad-line components and outflows. We define the narrow line component as the sum of all Gaussian components with $\rm FWHM < 1200\ km\;s^{-1}$, and the broad line component as the sum of all Gaussian parts with $\rm FWHM \geq 1200\ km\;s^{-1}$. Figure~\ref{fig:all_sdss_spec} and ~\ref{fig:all_gtc_spec} show the best fitting results of SDSS and GTC spectra of all the quasars, respectively.

\subsection{Light Curve Analysis}

\subsubsection{Data Ensemble}
We collect the optical $g, r, i$ bands, and mid-infrared (mid-IR) W1 and W2 bands light curves from various catalogs.

The ZTF instrument is installed on the 48-inch Samuel Oschin Telescope at the Palomar Observatory. It has a wide field-of-view (47 $\rm deg^2$) and multiple broad band filters. The ZTF survey scans the Northern sky at a cadence of around two days. We retrieve the $g, r$, and $i$ band light curves from ZTF DR12 (released on July 7, 2022) from the NASA/IPAC Infrared Science Archive \citep[IRSA,][]{https://doi.org/10.26131/irsa539}. The total observation duration is from March 2018 to May 2022, i.e. MJD from about 58200 to 59700. 

Near-Earth object Wide-field Infrared Survey Explorer  \citep[NEOWISE,][]{2010AJ....140.1868W, 2014ApJ...792...30M} is a repurposed survey after the WISE mission. It provides all-sky photometric monitoring at 3.4 $\mu m$ (W1) and 4.6 $\mu m$ (W2). We retrieve the photometry from the NASA/IPAC IRSA public data archive \citep{https://doi.org/10.26131/irsa144}. The cadence of the NEOWISE light curve is about 200 days, and the light curve spans a time range from about MJD = 56600 to about 59400. On every single epoch, NEOWISE has several exposures, we adopt the median value and the standard deviations as its uncertainties. The NEOWISE photometry is converted from Vega magnitude to AB magnitude by $m_{\rm AB} = m_{\rm Vega} + \Delta m$, where $\Delta m$ is 2.699 for W1 band and 3.339 for W2 band. We also retrieve the optical photometry in previous epochs from SDSS DR15 \citep{2019ApJS..240...23A}, PTF  \citep{2009PASP..121.1395L, https://doi.org/10.26131/irsa155}, Pan-STARRS1 DR2 \citep{2020ApJS..251....7F}, and Gaia DR3 \citep{2016A&A...595A...1G, 2022arXiv220800211G}. Notably, there is still a significant time gap in these early stages. We also try to retrieve the UV, X-ray, $\gamma$-ray, or radio detection of these quasars. Unfortunately, the detection of these bands is rare, most of them have one data point or only the upper limit. Figure~\ref{fig:all_lightCurve} shows the optical $r$-band light curves of the other ten quasars. 


\subsubsection{Dust Reverberation Analysis}

To further study the torus properties of these quasars, we analyze the dust reverberation signals by measuring the mid-IR-optical time lag. The optical light curve is from ZTF $r$ band and the mid-IR light curves are from WISE W1 and W2 bands. The time lag is measured by the traditional method, i.e., interpolated cross-correlation function \citep[ICCF,][]{1998PASP..110..660P}. ICCF is the most common method to measure the time lag between the two light curves. It is an empirical method to recover lost information of variability by adopting linear interpolation to the light curve and calculating the cross-correlation function to find the time lag. The Pearson correlation coefficient ($r$) is used to evaluate the significance of the time lag measurement. A larger $r$ suggests a better correlation between the two light curves \citep{2017ApJ...851...21G, 2020ApJ...900...58Y}. In this work, we adopt the \texttt{Python} script \texttt{PyCCF} \citep{2018ascl.soft05032S} to perform the ICCF calculations.


\section{Results}


\begin{figure}
   \centering
   \includegraphics[width=\columnwidth]{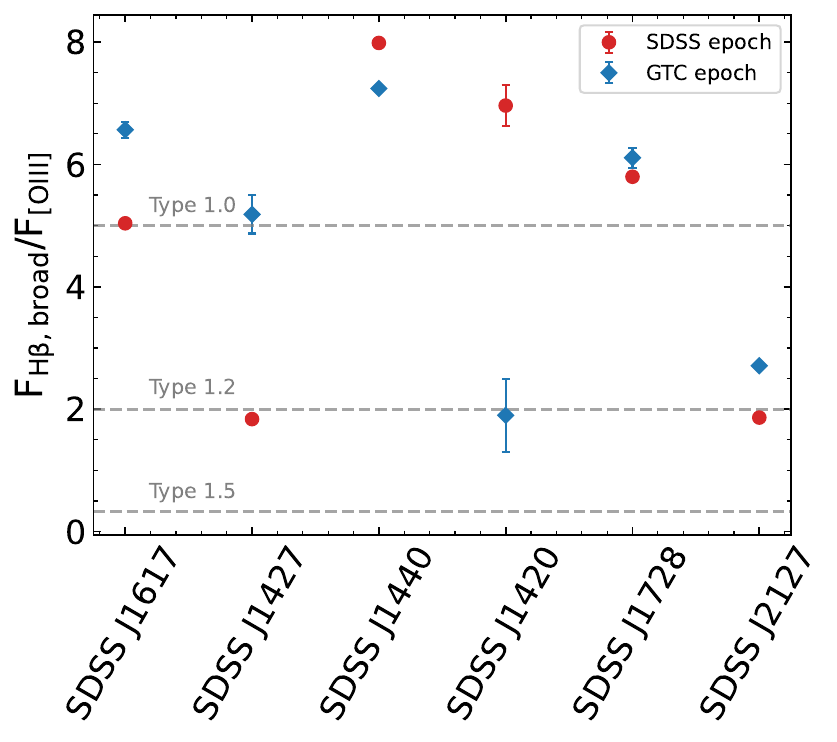}
      \caption{Spectral type of six quasars. We measure the spectral type of our quasars by the flux ratio between the broad line H$\beta$ and [\ion{O}{iii}] \citep{1992MNRAS.257..677W}. The red dots mark the SDSS epoch, while the sky-blue diamond represents the GTC epoch. These six quasars are selected with available H$\beta$ and [\ion{O}{iii}] emission line measurements.}
         \label{fig:specType}
   \end{figure}

\begin{figure}
   \centering
   \includegraphics[width=\columnwidth]{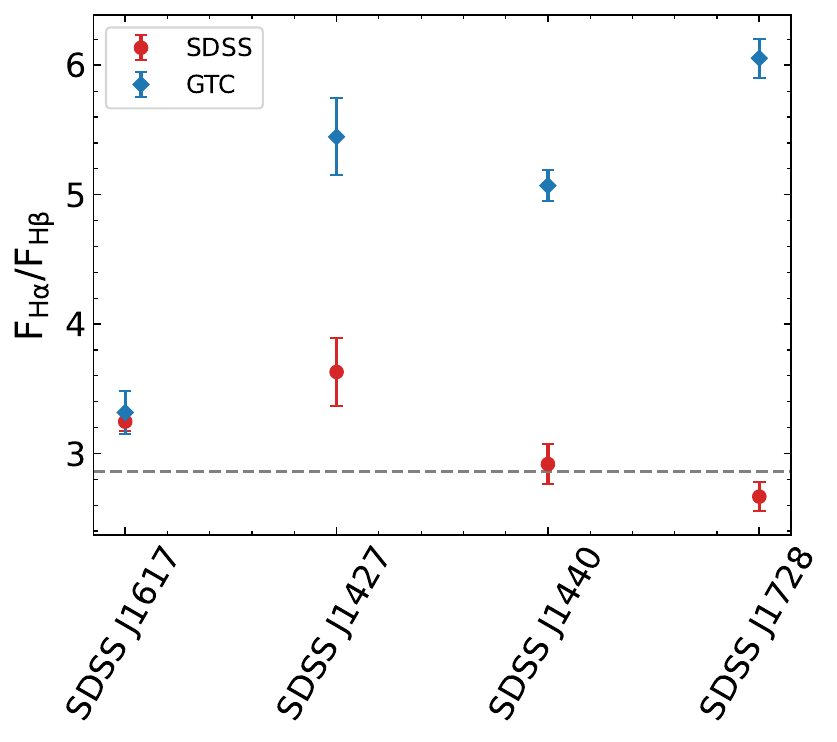}
      \caption{Balmer decrement of four quasars with both H$\alpha$ and H$\beta$ emission line measurements. The markers are the same as the Figure~\ref{fig:specType}. The grey dashed line represents the Case B recombination, where the ratio of $\rm H \alpha$ and $\rm H \beta$ is 2.86.}
         \label{fig:BalmerDecrement}
   \end{figure}

\begin{figure*}
   \resizebox{\hsize}{!}
             {\includegraphics[width=\textwidth]{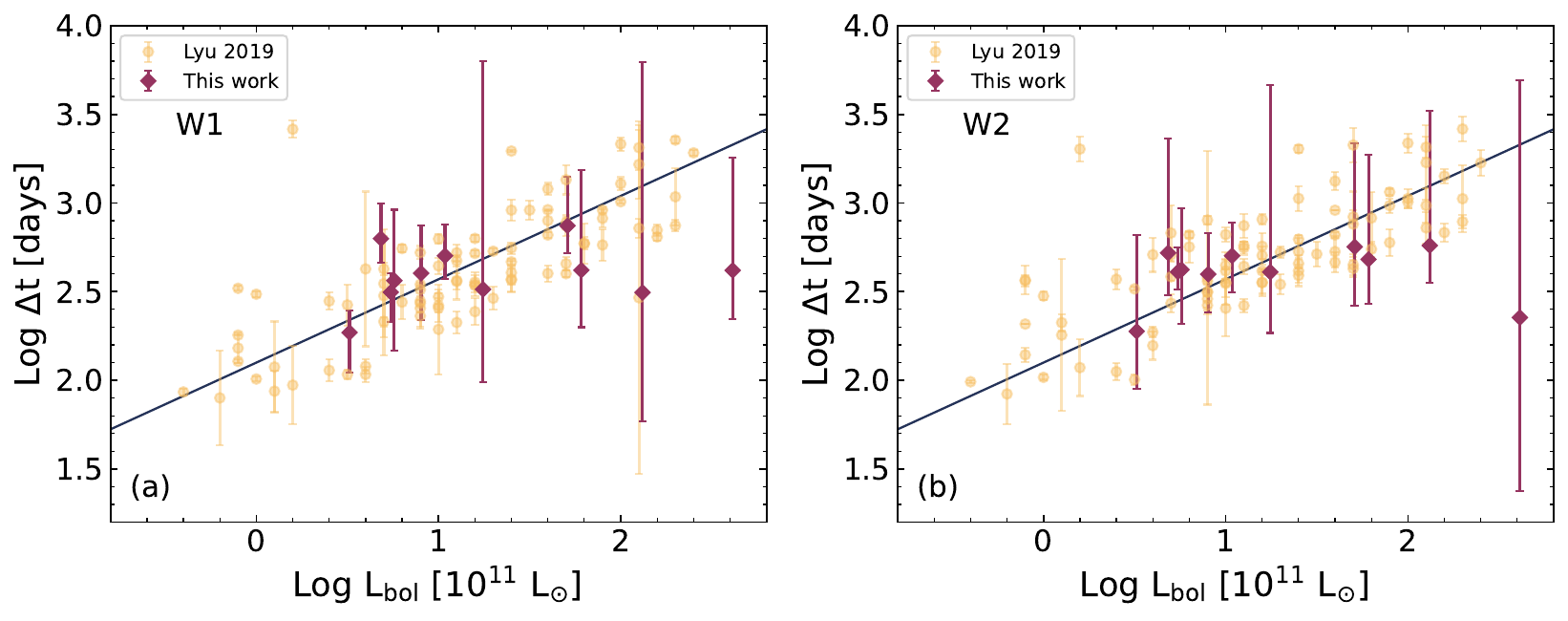}}
      \caption{mid-IR to optical light curve time lag as a function of the AGN bolometric luminosity. We compare the measured time lag of our quasars with the PG quasars \citep{2019ApJ...886...33L}, both the light-orange dots and the best-fit relation are from  \citet{2019ApJ...886...33L}, and the dark-purple diamond shows our results. We measure the mid-IR to optical light curve time lag through the traditional ICCF method. The subplots (a) and (b) show the time lag between WISE W1 (3.4 $\mu m$),  WISE W2 (4.6 $\mu m$), and ZTF $r$ band, respectively.}
         \label{fig:timeLag}
   \end{figure*}

\begin{figure*}
   \resizebox{\hsize}{!}
             {\includegraphics[width=\textwidth]{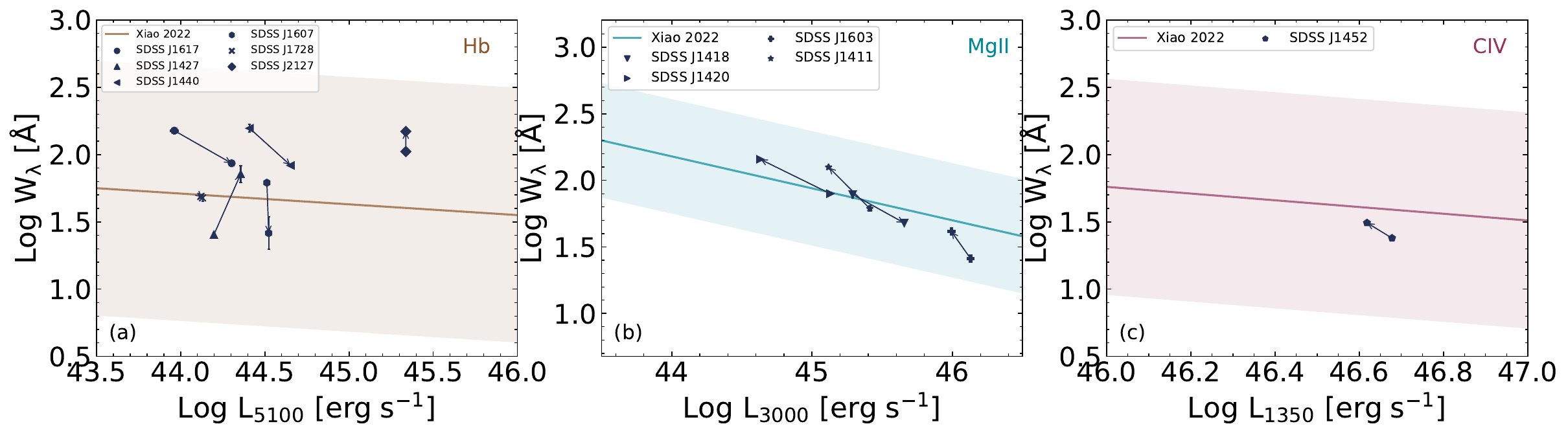}}
      \caption{Baldwin effect of our quasars. The different broad lines and continuum luminosity are utilized in different redshift, (a) $z < 0.7$: $\rm H \beta $ and $\rm L_{5100}$, (b) $0.7 < z < 1.9$: \ion{Mg}{ii} and $\rm L_{3000}$, (c) $z > 1.9$: \ion{C}{iv} and $\rm L_{1350}$. For every quasar, we use a vector to exhibit its variation from the SDSS epoch ($\sim$ MJD 53000) to the GTC epoch ($\sim$ MJD 59000). In each panel, the background best-fit relation and its 1 $\sigma$ scatter shadow are from \citet{2022ApJ...936..146X}. }
         \label{fig:baldwineffect}
   \end{figure*}

\begin{figure}
   \centering
   \includegraphics[width=\columnwidth]{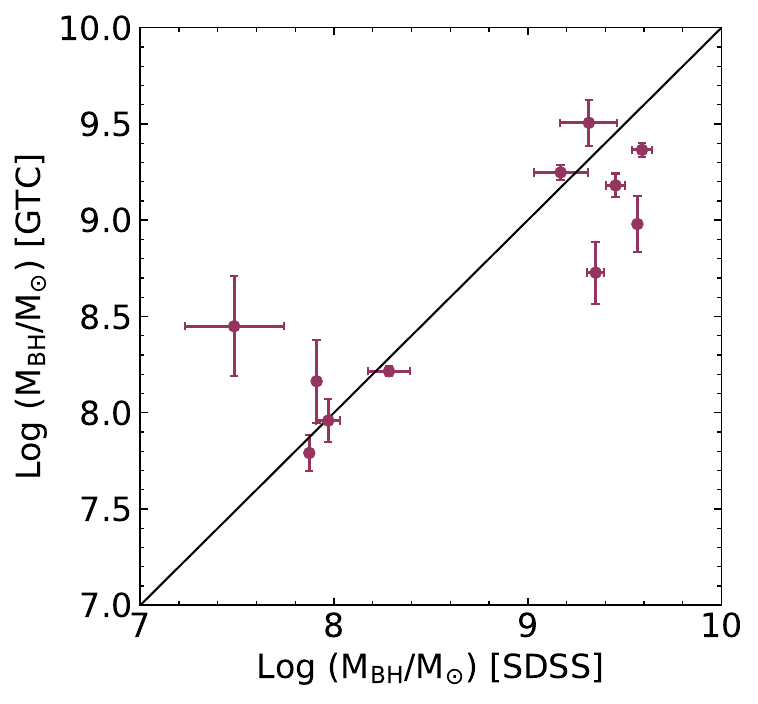}
      \caption{The comparison of SMBH masses between the SDSS and GTC epochs, where the black line indicates a 1:1 ratio. The SMBH masses are estimated using different broad emission lines at different redshift bins: $\rm H\beta$ for $z<0.7$, \ion{Mg}{ii} for $ 0.7 \leq z <1.9$, and \ion{C}{iv} for $z \geq 1.9$.}
         \label{fig:bhmass}
   \end{figure}
   
\begin{figure}
   \centering
   \includegraphics[width=\columnwidth]{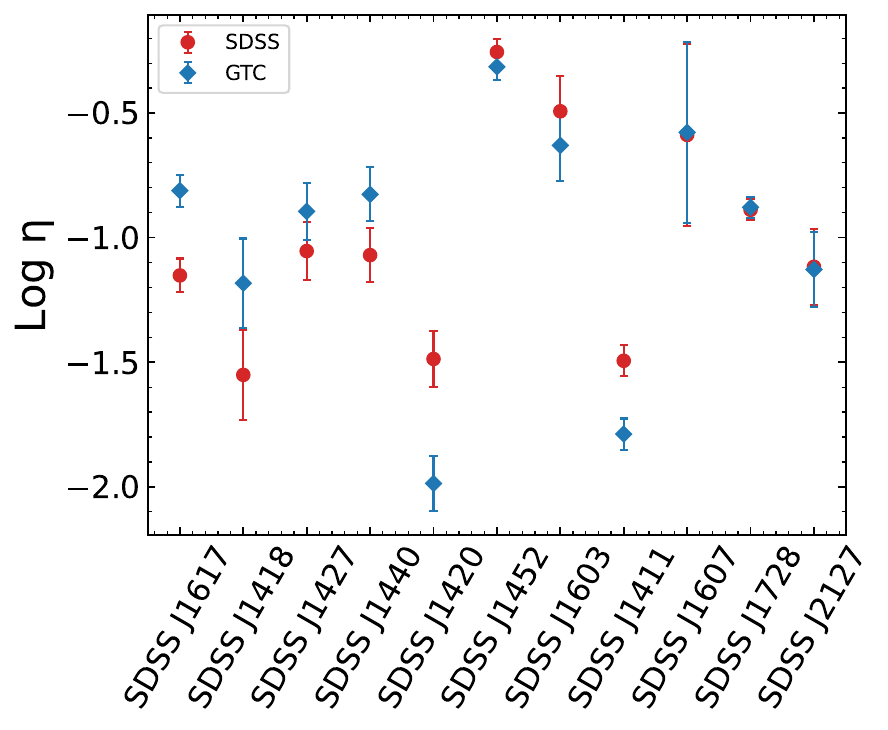}
      \caption{The Eddington ratios comparison between SDSS and GTC epoch of our quasars. The markers are the same as Figure~\ref{fig:specType}.}
         \label{fig:eddingtonRatio}
   \end{figure}


\subsection{Spectral Type}

We investigate the spectral type of our quasars at two epochs to search for potential CL-AGN candidates. The spectral types of our quasars are classified into type 1, 1.2, and 1.5 according to the ratio between the broad $\rm H \beta$ line flux and the total flux of [\ion{O}{iii}] \citep{1992MNRAS.257..677W}. Figure~\ref{fig:specType} shows the spectral types of six quasars with available H$\beta$ and [\ion{O}{iii}] emission lines due to the redshift limitation. There are two quasars that may be the CL-AGN candidates: SDSS J1427+2930 changes its spectral type from type 1.2 to type 1.0, and SDSS J1420+3757 changes from type 1.0 to type 1.2. Other quasars do not show a significant change in the spectral type in this classification criteria.

\subsection{Broad-line Balmer Decrement}

The foreground obscuration variation can be simply described as dust clouds moving into (out of) the line of sight and making the observed luminosity increase (decrease). To cause rapid variability, the dust clouds should be close to the SMBHs that may be recently captured by the SMBH from the dust torus \citep{2017NatAs...1..775W}. Such a scenario can produce unique features in its spectra: the equivalent width of broad lines ($\rm H \beta$, \ion{Mg}{ii}, and \ion{C}{iv}) should be consistent in SDSS and GTC spectra, the continuum variation should follow the attenuation change. 

The flux ratio between Balmer lines, i.e. Balmer decrement, is widely adopted to describe the nebular dust attenuation state \citep{2002A&A...383..801B, 2012MNRAS.419.1402G}. We utilize the flux ratio between the broad line H$\alpha$ and H$\beta$ to study the dust attenuation in front of the broad line region (BLR) near the SMBH. Figure~\ref{fig:BalmerDecrement} shows the broad-line Balmer decrement of four quasars with well-observed H$\alpha$ and H$\beta$ emissions. The results indicate that three of four quasars show large changes in the broad-line Balmer decrement. The remaining quasar shows no obvious Balmer decrement change between the two epochs. However, the Balmer decrement could also be the result of the change in the ionization state. 

To further point out which mechanism causes such a change in our quasars, we investigate the torus size-luminosity relation \citep{2006ApJ...639...46S} for the obscuration explanation and Baldwin effect \citep{1977ApJ...214..679B, 1992AJ....103.1084P} for the ionization mechanism. On the one hand, we draw the time lag between the optical $r$ band and the mid-IR W1 and W2 band light curves in Figure~\ref{fig:timeLag} (a) and (b), respectively. In both panels, the time lag is compared to  PG quasars \citep{2019ApJ...886...33L} that have an ordinary torus. The time lags of our quasars are consistent with the best-fit linear relation of PG quasars within the uncertainties, indicating our quasars are similar in the torus size to PG quasars though the errors are large. On the other hand, the equivalent width varies with continuum luminosity between two epochs are shown in Figure~\ref{fig:baldwineffect} (a) - (c). We separate 11 quasars into 3 subsamples according to their redshift. For $z < 0.7$, we use the $\rm H \beta $ and $\rm L_{5100}$; for $0.7 < z < 1.9$, we use the \ion{Mg}{ii} and $\rm L_{3000}$; for $z > 1.9$, we use the \ion{C}{iv} and $\rm L_{1350}$. In each panel, the variation between the two epochs is drawn as a vector and then compared with the best-fit relation of various quasars  \citep{2022ApJ...936..146X}. The vectors are consistent with the relationship within 1 $\sigma$ scatter, while the directions of vectors are diverse.

\subsection{Accretion Rate}

Another possible origin of the rapid luminosity change is the enhancement or diminishing of the accretion rate. We measure the virial mass of the SMBH at the two epochs and adopt the average value of the final SMBH mass \citep{2011ApJS..194...45S}. The different broad emission lines are used in different redshift bins: $\rm H\beta$ for $z<0.7$, \ion{Mg}{ii} for $ 0.7 \leq z <1.9$, and \ion{C}{iv} for $z \geq 1.9$. The uncertainties of the SMBH masses are estimated from the flux and FWHM errors of each broad line which are fitted by the \texttt{PyQSOFiT}. Fig~\ref{fig:bhmass} compares SMBH masses between the two epochs. The bolometric luminosity is estimated by the $L_{5100}$, $L_{3000}$, and $L_{1350}$ for above different redshift bins. After the SMBH mass and bolometric luminosity measurements, the Eddington ratio at two epochs is defined as $\eta \equiv L_{\rm bol}/L_{\rm Edd}$, where $L_{\rm Edd} = 1.3 \times 10^{38} (M_{\rm BH}/{\rm M_{\odot}})\ {\rm erg\;s^{-1}}$. Figure~\ref{fig:eddingtonRatio} indicates that most of our quasars show a slight change in the accretion rate, and several quasars show significant differences in the Eddington ratio between the two epochs, including SDSS J1420+3757, SDSS J1427+2930.


\section{Discussion}

\subsection{Possible Origins of the Flare/Eclipse-like Variability}

The two epochs' spectra and multi-band light curves provide rich information for understanding the origin of the optical variability of these quasars. It might be the dust clouds moving through our line of sight cause the observed flux rise or fall, which is consistent with the changes in the Balmer decrement. However, the Balmer decrement change is not necessarily caused by the obscuration state change, especially in low-luminosity AGNs and CL-AGNs \citep{2023arXiv230409435W}. To further clarify which mechanism affects the Balmer decrement, we not only measure the time lag between MIR and optical light curves but also the Baldwin effect. From the time lag measurements, we obtain a normal torus size of our quasars, which rules out the scenario that the SMBH captures dust clouds from the torus and changes the broad line Balmer decrement. In this case, such a scenario should cause a much longer timescale luminosity change, more than half a year at least. Besides, the mid-IR echo also supports the accretion rate change scheme \citep{2017ApJ...846L...7S}. On the other hand, recent studies rise observational evidence of the polar region dust that the extended dust structures align with the narrow line region (NLR) \citep{2022Univ....8..304L}. Despite the torus, the polar region dust clouds also can travel through the line of sight and cause rapid luminosity change. However, the clouds from the extended polar region dust will cause a much longer time scale for the AGN variability, rather than the rapid luminosity change. 

The other mechanism to cause the Balmer decrement change is the change of ionization state. We conclude that the broad line variations of our quasars follow the Baldwin effect, which may be a clue for the ionization state changes although the origin of the Baldwin effect is still under debate. Lots of mechanisms are proposed as the underlying physics of the Baldwin effect \citep{1984ApJ...278..558M, 1985MNRAS.216...63N, 1990ApJ...357..338K, 1992AJ....103.1084P, 2002ApJ...581..912D, 2012MNRAS.427.2881B}. The most accepted explanation is that the continuum becomes softer with the AGN bolometric luminosity rise so that the high bolometric luminosity AGN emits fewer high energy photons to ionize the broad emission lines \citep{1993ApJ...415..517Z}. Besides, the Eddington ratio change is also a possible origin of the Baldwin effect \citep{2004ApJ...617..171B, 2012MNRAS.427.2881B}. As a result, the Baldwin effect can be circumstantial evidence of the ionization state change of the gas material around the SMBH that is driven by the accretion rate changes. Figure~\ref{fig:eddingtonRatio} shows that our quasars are located at the low Eddington ratio regime, and almost all represent the changes in the Eddington ratio. These features are consistent with those known extreme AGN variabilities and CL-AGNs that are triggered by accretion rate changes \citep{2018ApJ...854..160R, 2022ApJ...933..180G}. Consequently, it is preferred that the accretion rate variation accounts for the rapid AGN variability in our sample, rather than the obscuration changes. 

\subsection{Two CL-AGN Candidates}
\subsubsection{SDSS J1427+2930}

\begin{figure*}
   \resizebox{\hsize}{!}
             {\includegraphics[width=\textwidth]{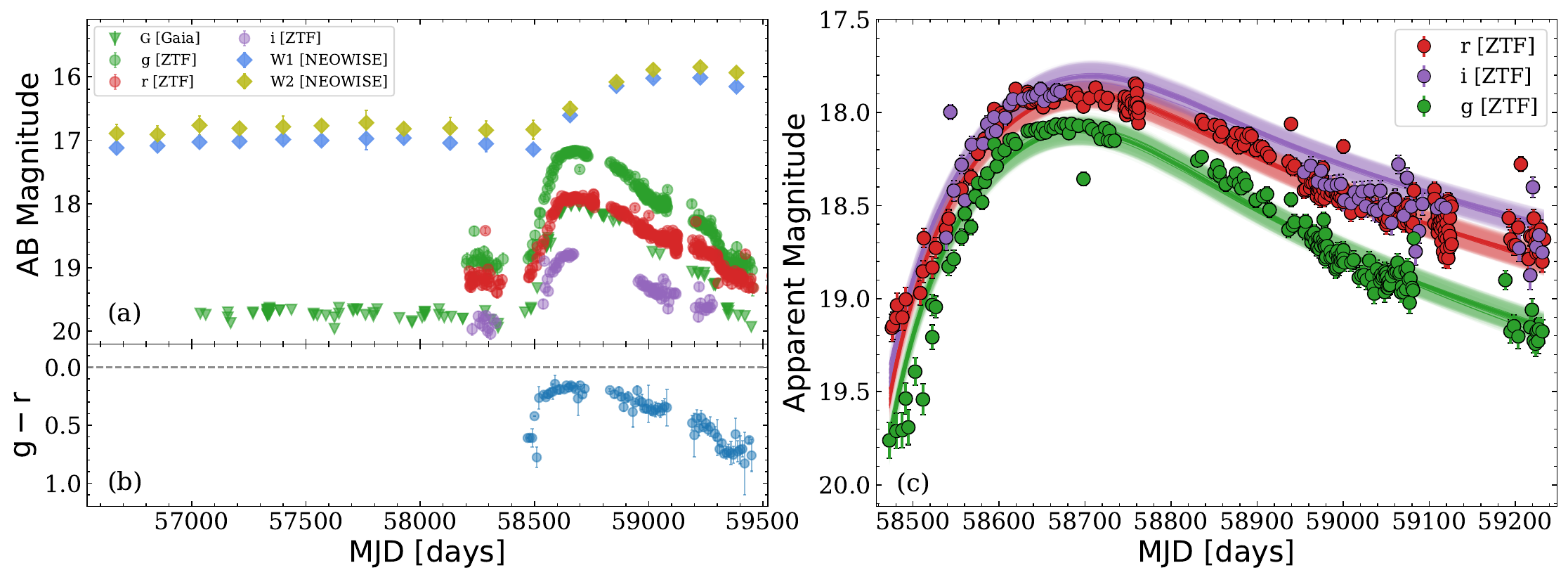}}
      \caption{The multi-band light curve and the fitting results of SDSS J1427+2930. (a): multi-band light curve. The green anti-triangles are the Gaia photometry in the $g$ band. For the optical light curve from ZTF, $g$, $r$, and $i$ bands are marked by the green, red, and purple circles, while the yellow and blue diamonds represent the W1 and W2 bands mid-IR light curve from neoWISE. (b): $g-r$ color estimated from the ZTF light curve. (c): best fitting results from \texttt{MOSFiT}. The 3 bands optical light curve is modeled by the TDE model from \citet{2019ApJ...872..151M}, and gives a best-fit result that an SMBH with a mass of $\sim 10^{8.37}\; {\rm M_{\odot}}$ destroy a giant star of $ 14.20^{+1.23}_{-0.40}\;\rm M_{\odot}$.}
         \label{fig:1427lc}
   \end{figure*}

\begin{figure}
   \centering
   \includegraphics[width=\columnwidth]{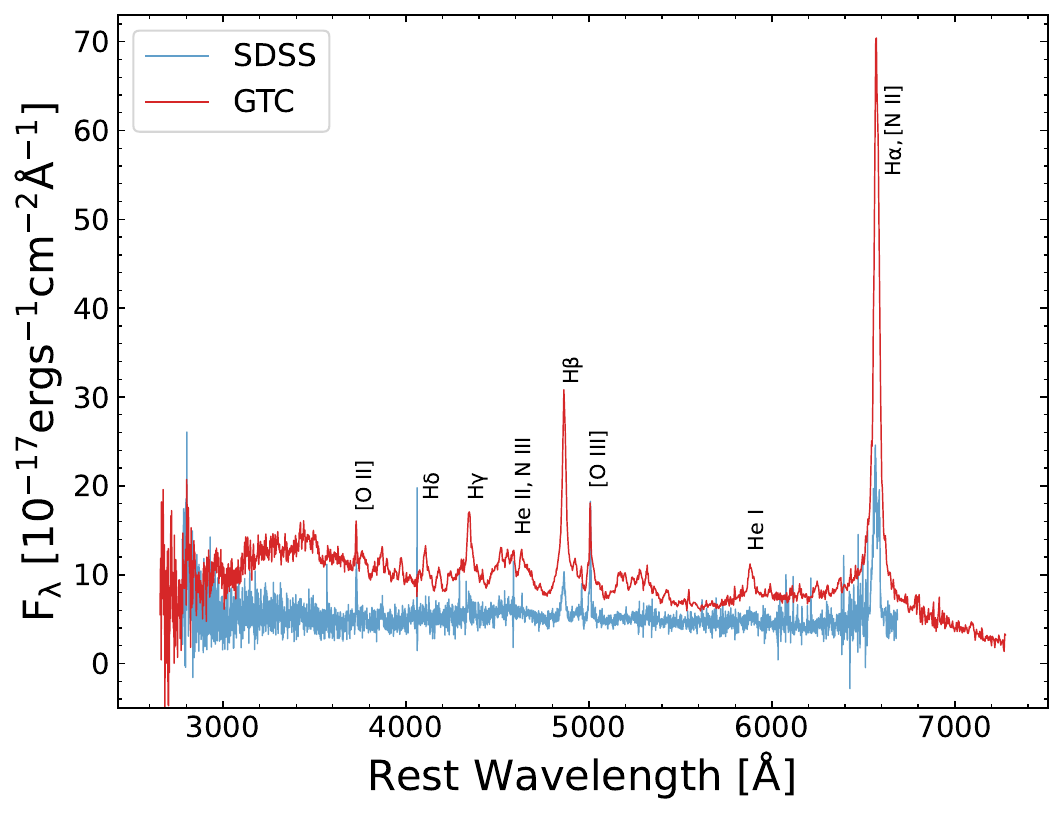}
      \caption{The spectra between two epochs of SDSS J1427+2930. The blue and red spectra represent the SDSS epoch and the GTC epoch. The emission lines are marked at their rest-frame wavelength.}
         \label{fig:1427spec}
   \end{figure}


Figure~\ref{fig:specType} indicates the quasar SDSS J1427+2930 is a possible CL-AGN that appears to have transitioned from type 1.2 to type 1 between MJD 53917 and MJD 59054. Years ago, it was reported as an SMBH-related flare (AT2019brs) by Folded Low
Order whYte-pupil Double-dispersed Spectrograph North (FLOYDS-N) follow-up \citep{2019TNSTR.366....1D, 2019TNSAN...4....1A}. Later, \citet{2021ApJ...920...56F} further studied its multi-epoch spectra and estimated its SMBH mass of $\sim 10^{8.2}\;{\rm M_{\odot}}$ from the $V$-band absolute magnitude of its host galaxy. The star captured by such a massive SMBH should fall into the SMBH directly without tidal disruption. For a solar mass star and a non-spin BH, the SMBH mass limit of TDE occurrence is $\sim 10^8\ {\rm M_{\odot}}$ which is called the Hill limit \citep{1975Natur.254..295H}. Meanwhile, Figure~\ref{fig:1427lc} (b) shows the $g-r$ color which shows a non-flat evolution. Therefore, \citet{2021ApJ...920...56F} ruled out the TDE scenario due to the large SMBH mass and the non-flat color evolution. They classified it as an AGN flare that is related to the enhanced SMBH accretion proposed recently, according to the strong Bowen fluorescence emissions and the long-evolved light curve \citep{2019ApJ...883...31F, 2019NatAs...3..242T}.

It is not impossible for TDE to occur in such a massive SMBH, for example, a spin SMBH or a giant star \citep{2016NatAs...1E...2L, 2019MNRAS.487.4790G, 2023ApJ...948L..19S}. Here, we adopt the Python program \texttt{MOSFiT} \footnote{\url{https://mosfit.readthedocs.io/en/latest/}} \citep{2018ApJS..236....6G} to fit the optical light curve of SDSS J1427+2930 with a TDE model \citep{2019ApJ...872..151M}. This program models the multi-band light curve of transients with various triggering mechanisms by using a Markov Chain Monte Carlo sampler. We utilize the default parameter settings with a few modifications, i.e. larger ranges of SMBH and stellar masses. The best-fit model gives a result that a SMBH with ${\rm log(M_{BH}/{\rm M_{\odot}}) = 8.37\pm0.02}$ tidally destroys a giant star of $ 14.20^{+1.23}_{-0.40}\;\rm M_{\odot}$. The disruption factor is quantified by the parameter $b = 0.88 \pm 0.01$, which is the scaled impact parameter $\beta$. The parameter $b$ can vary from 0 to 1, which corresponds to the minimum disruption and full disruption. \citet{2019ApJ...872..151M} estimated the systematic uncertainties of the SMBH and stellar masses from the TDE model to 0.2 dex and 0.66 dex, respectively. Figure~\ref{fig:1427lc} (c) shows the best-fit results from \texttt{MOSFiT}. Figure~\ref{fig:1427lc} (a) shows the multi-band light curves, which exhibit an echo in the mid-IR light curve. The MIR echos are seen in some TDE candidates \citep{2016ApJ...828L..14J, 2021ARA&A..59...21G} which indicates that SDSS J1427+2930 might be a TDE in AGN. Besides, \citet{2021ApJ...920...56F} observed a broad \ion{He}{ii} emission (${\rm FWHM} \sim 6000 \; {\rm km\;s^{-1}}$) at 133 days after the peak luminosity. As shown in Figure~\ref{fig:1427spec}, the \ion{He}{ii} emission also exists although the GTC spectra are observed at around 600 days after the peak, which indicates a possible TDE scenario.  

These complex features in multi-band light curves and multi-epoch spectra can be triggered by not only the AGN flare but also the TDE scenario. Recently, \citet{2021SSRv..217...54Z} summarized how we could distinguish the TDE from other transients including the extreme AGN flare. For the two phenomena with highly similar observed features, heaps of features are proposed to find the TDE candidates in AGNs. In this case, even the well-studied candidates could not meet all the features they listed, such as the ASASSN-14li, ASASSN-15oi, etc. SDSS J1427+2930 does meet several features, 1) a steep bright rise in optical flux, 2) a bluer $g-r$ color around the luminosity peak, 3) weaken [\ion{O}{iii}] emission line, 4) a broad \ion{He}{ii} emissions. However, in certain conditions, it is possible that a TDE candidate could not satisfy all the features, or another transient might meet all these features but not a TDE. It seems that the features in photometry and spectroscopy observed to date could not support us in distinguishing the TDE from AGNs. In the case of SDSS J1427+2930, it is also difficult to certify its origin. In addition, the micro-TDE is proposed to explain the unique light curve \citep{2022ApJ...933L..28Y}. The micro-TDE describes a stellar-mass black hole destroying a star in the accretion disk of the central SMBH, which might produce a similar light curve shape.

\subsubsection{SDSS J1420+3757}

\begin{figure*}
   \resizebox{\hsize}{!}
             {\includegraphics[width=\textwidth]{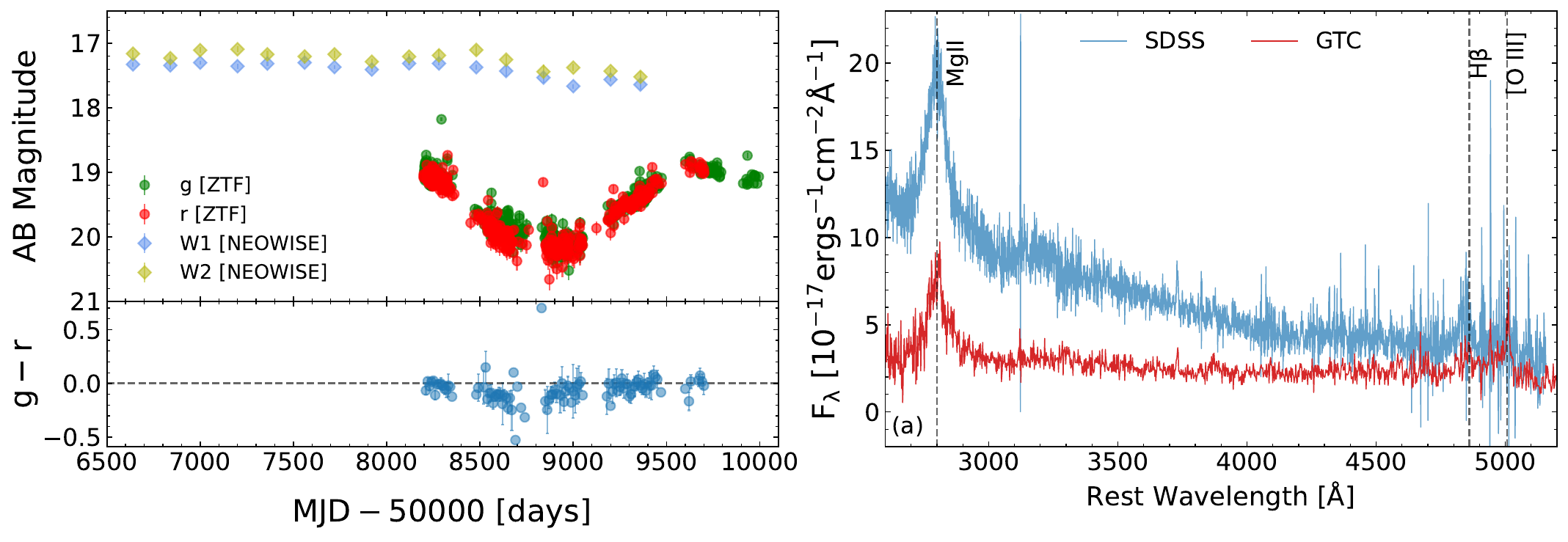}}
      \caption{The multi-band photometry and multi-epoch spectra of SDSS J1420+3757. Left: multi-band light curves and the 'g-r' color evolution. The green and red circles mark the ZTF $g$ and $r$ band, respectively. The yellow and blue diamonds represent the median mid-IR W1 and W2 band light curves from NEOWISE. The $g-r$ color is estimated from the ZTF light curve. Right: multi-epoch spectra. The blue and red spectra represent the SDSS epoch and the GTC epoch spectra, respectively. The \ion{Mg}{ii}, H$\beta$, and [\ion{O}{iii}] emission lines are marked by the vertical black lines.}
         \label{fig:J1420_results}
   \end{figure*}

As shown in Figure~\ref{fig:specType}, we recognize SDSS J1420+3757 as a CL-AGN candidate by its spectral type change. Under the classification criterion in \citet{1992MNRAS.257..677W}, it changes from Type 1 to Type 1.2. Figure~\ref{fig:J1420_results} shows the two epochs' spectra and the $r$-band light curve. This quasar has been undergoing a dramatic luminosity change in the recent few years which dims its brightness over 1 magnitude and then re-bright. It meets some features which are similar to changing-state AGNs (CS-AGNs) as triggered by the accretion state changes. A direct feature is that the Eddington ratio decreases by about 0.5 dex during the two epochs. Another property is the changes in \ion{Mg}{ii} emission line, whose flux also decreases with the decreasing continuum brightness. Besides, Figure~\ref{fig:baldwineffect} (b) shows that the change in \ion{Mg}{ii} EW follows the expectation by the Baldwin effect nicely. These features are seen in the CL-AGNs that are selected by the \ion{Mg}{ii} variability \citep{2019ApJ...883L..44G, 2020ApJ...888...58G}. Therefore, the huge change in the accretion rate more likely accounts for its rapid variability and spectral type change.


\section{Conclusions}

With the high-cadence monitor onboard ZTF, we have identified a rare sample of flare/eclipse-like AGN variability. Combining the multi-epoch spectra and multi-band light curves, we find the rapid luminosity change in these quasars is driven by their accretion rate variation. Our main conclusions are as follows.

\begin{itemize}

    \item In our sample, four out of the 11 quasars have reliable Balmer decrement measurements, and three quasars show significant changes in the broad-line Balmer decrement. Notably, three out of the 11 quasars show no obvious variation in the Eddington ratio. All 11 quasars in our study exhibit characteristics consistent with a standard torus size and adhere to the Baldwin effect.

    \item In line with the Baldwin effect, it is more plausible that the detected variations in the Balmer decrement are attributed to shifts in the ionization state. Additionally, the presence of the Baldwin effect may indicate that the change in accretion rate plays a substantial role in these flare/eclipse-like AGN variabilities.


    \item Following the spectral type diagnosis, we find two CL-AGN candidates, SDSS J1420+3757 and SDSS J1427+2930. The former changes its type from type 1 to type 1.2, while the latter changes from type 1.2 to type 1 between the SDSS and GTC periods.

    \item For SDSS J1420+3757, the broad-line variation is aligned with its dimm luminosity, and the Eddington ratio also shows a significant change. Therefore, we conclude that its change in the accretion rate causes rapid variability in the optical light curve. 

    \item The situation in SDSS J1427+2930 is quite complex. It shows a unique shape in the light curve and plenty of features in the spectra. Not only the special AGN flare but also the TDE can reproduce these features. However, the features observed to date are still impossible to distinguish the TDE from the AGN. 
    
\end{itemize}


\section*{Acknowledgements}

Z.Z. and Y.S. acknowledge the support from the National Key R\&D Program of China No. 2022YFF0503401, the National Natural Science Foundation of China (NSFC grants 11825302, 12141301, 12121003, 12333002). SJ and HD acknowledge financial support from the Spanish Ministry of Science, Innovation and Universities (MICIU) under grant AYA2017-84061-P, co-financed by FEDER (European Regional Development Funds). HD acknowledges financial support from the Agencia Estatal de Investigación del Ministerio de Ciencia e Innovación (AEI-MCINN) under grant (La evolución de los cúmulos de galaxias desde el amanecer hasta el mediodía cósmico) with reference (PID2019-105776GB-I00/DOI:10.13039/501100011033). SJ is supported by the European Union's Horizon Europe research and innovation program under the Marie Sk\l{}odowska-Curie grant agreement No. 101060888. X.L.Y. acknowledges the grant from the National Natural Science Foundation of China (NSFC grants 12303012), Yunnan Fundamental Research Projects (No.202301AT070242).

\section*{Data Availability}

The data underlying this article will be shared on reasonable request to the corresponding author.



\bibliographystyle{mnras}
\bibliography{rapidQSOvar} 



\bsp	
\label{lastpage}
\end{document}